\documentclass[prb,reprint,aps,superscriptaddress,floatfix]{revtex4-2}
\usepackage{graphicx}
\usepackage{color}
\usepackage{amsmath}
\usepackage{natmove}
\usepackage{natbib}
\usepackage{hyperref}

\begin{document}
\title{Effect of the electronic charge gap on LO bond-stretching phonons in undoped La$_\text{2}$CuO$_\text{4}$ calculated using LDA+U}

\author{Tyler C. Sterling}
\email{ty.sterling@colorado.edu}
\affiliation{Department of Physics, University of Colorado at Boulder, Boulder, Colorado 80309, USA}

\author{Dmitry Reznik}
\affiliation{Department of Physics, University of Colorado at Boulder, Boulder, Colorado 80309, USA}
\affiliation{Center for Experiments on Quantum Materials, University of Colorado at Boulder, Boulder, Colorado 80309, USA}

\date{\today}

\begin{abstract}
Typical density-functional theory calculations that wrongly predict undoped cuprates to be metallic also predict Cu-O half- and full-breathing phonon energies that are significantly softer than observed, presumably because of weak on-site Coulomb repulsion on the Cu 3$d$ orbitals. We used DFT+U calculations with antiferromagnetic supercells of La$_\text{2}$CuO$_\text{4}$ to establish correlation between the on-site repulsion strength, tuned via adjusting the value of U, and phonon dispersions. We find that breathing and half-breathing phonons reach experimental values when U is tuned to obtain the correct optical gap and magnetic moments. We demonstrate that using distorted supercells within DFT+U is a promising framework to model phonons in undoped cuprates and other perovskite oxides with complex, interrelated structural and electronic degrees of freedom.
\end{abstract}

\maketitle

\section{Introduction}

Phonons in cuprates have been studied mostly in doped phases in an effort to explain superconductivity and charge-density wave physics \cite{McQueeney99,park2014evidence,reznik2008temperature,ahmadova2020phonon,giustino2008small,lebert2020doping,le_tacon_inelastic_2014,pintschovius2005electron,reznik2012phonon}. However, it is becoming increasingly apparent that phonons in \emph{undoped} cuprates merit further investigation. For example, La$_\text{2}$CuO$_\text{4}$ hosts interesting phonon physics even in the undoped phase. There are energy lowering structural distortions \cite{boni1988lattice,birgeneau1987soft,billinge1996probing,bozin2015reconciliation,sapkota2021reinvestigation}, spin-orbit induced magnetic behavior \cite{shekhtman1992moriya,thio1994weak}, and most recently a large thermal Hall effect has been observed and attributed to phonons \cite{grissonnanche2019giant,grissonnanche2020chiral}. Unfortunately, the overwhelming majority of previous phonon calculations for undoped cuprates used density-functional theory (DFT) in either the local density approximation (LDA) or the generalized gradient approximation (GGA) \cite{cohen1990first,wang1999first,krakauer1993large,singh1996phonons,lebert2020doping,miao2018incommensurate,ahmadova2020phonon}.

Undoped cuprates are insulating and antiferromagnetic (AFM), but the LDA and GGA predict them to be metallic with either no magnetism or unrealistically small magnetic moments \cite{singh1991gradient,ambrosch1991local,giustino2008small,mattheiss1987electronic,yu1987electronically}. The disagreement is similar for the lattice dynamics. The LO bond-stretching phonons in undoped cuprates calculated from LDA or GGA do not agree with experiment but rather disperse steeply downward. Usually the bond-stretching dispersions calculated for undoped compounds fortuitously agree with experiments on overdoped compounds \cite{park2014evidence,pintschovius2006oxygen,pintschovius2005electron} and calculated dispersions are often presented along side experiments on overdoped materials \cite{ahmadova2020phonon,miao2018incommensurate,giustino2008small,lebert2020doping}. Most other branches are unaffected by doping and already match experiments in the LDA and GGA \cite{miao2018incommensurate,park2014evidence,pintschovius2006oxygen,pintschovius2005electron,ahmadova2020phonon,giustino2008small,lebert2020doping}. Calculating correct bond-stretching phonon dispersions of most undoped cuprates remains elusive.

The DFT+U method, which is an extension to DFT that includes an adjustable Hubbard-U like onsite repulsion on correlated orbitals (e.g, the Cu 3$d$ orbitals in cuprates), is known to predict reasonable gaps and moments for undoped cuprates and has been extensively applied to electronic-structure calculations across a wide range of doping \cite{zhang2007electron,pesant2011dft+,czyzyk1994local,anisimov1992spin,wei1994electronic,svane1992electronic,anisimov2004computation,elfimov2008theory,puggioni2009fermi,oh2011fermi}. However, aside from an earlier DFT+U calculation that succeeded in calculating correct energies of a few zone boundary phonons in CaCuO$_\text{2}$ \cite{zhang2007electron}, the phonon spectrum of DFT+U calculations in cuprates is mostly unexplored. 

\begin{figure}
\includegraphics[width=1\linewidth]{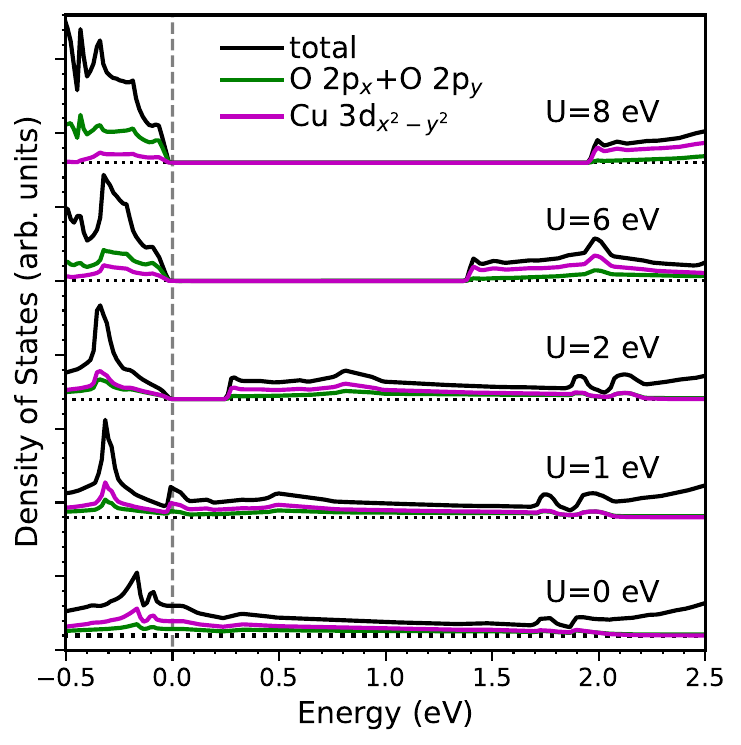}
        \caption{Electronic densities of states calculated using the U values indicated in the figure. The data are offset vertically for clarity. The colors indicate the orbitals the densities of states are projected onto. For U = 2, 6, and 8 eV, the electronic-structure is insulating with gaps $\approx$ 0.2, 1.4, and 2 eV, respectively. The magnetic moments for U = 1, 2, 6, and 8 eV are $\pm$ 0.21, 0.33, 0.53, and 0.61 $\mu_{B}$, respectively. For U=0 eV, the ground-state is metallic and nonmagnetic; for U=1 eV, it is metallic with small magnetic moments.}
\label{fig:dos}
\end{figure}

\begin{figure}
\includegraphics[width=1\linewidth]{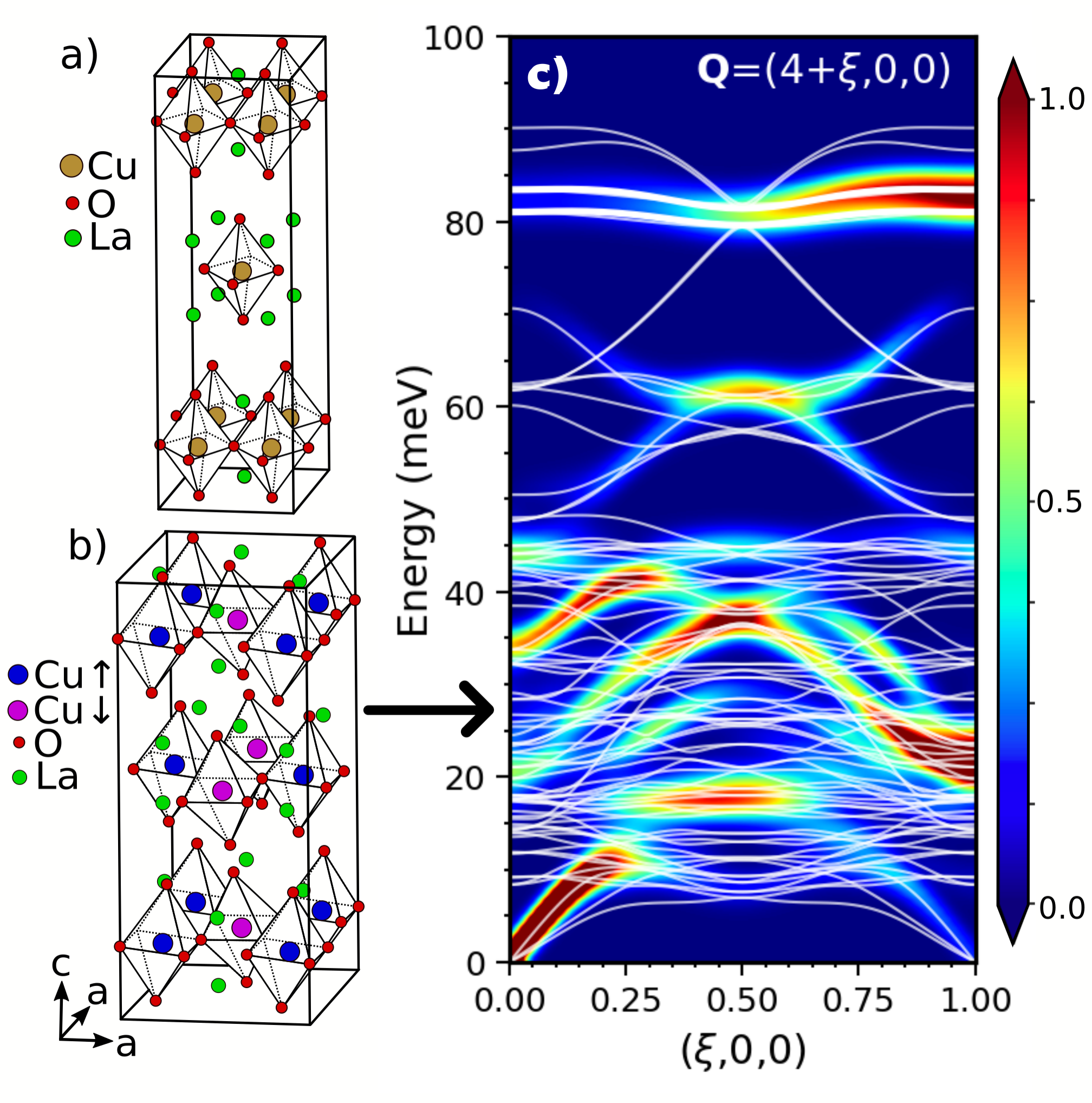}
        \caption{(a) The HTT cell of La$_\text{2}$CuO$_\text{4}$. (b) The LTT cell with correct AFM ordering and octahedral distortions. (c) Phonon dispersions and dynamic structure factors, S($\bf{Q}$,$\omega$), calculated in the LTT phase with U=8 eV. The $\bf{Q}$=$(4+\xi,0,0)$ zone is where the bond-stretching branch was measured \cite{pintschovius2006oxygen}. The structure factors are broadened with a Gaussian with 3 meV width. The white lines are the phonon dispersions in the first Brillouin zone.}
\label{fig:cells}
\end{figure}

Inspired by the apparent success for the electronic structure, we use the DFT+U method to investigate the interplay between the Cu 3$d$ on-site repulsion strength, tuned by varying U, and phonon dispersions. We already demonstrated spectacular agreement of the calculated acoustic phonons and nearby optic branches with experiment in another paper \cite{sapkota2021reinvestigation}. These-low lying branches are nearly independent of U and agree well even in plain GGA. This is not surprising since the low-energy phonons mainly involve motion of La and do not induce substantial charge redistribution around the Cu atoms. As such, dispersion of the low energy phonons are not a valuable metric for the accuracy of DFT+U. 

In this paper, we focus on the Cu-O bond-stretching phonons which are expected to depend strongly on the on-site repulsion and are therefore most suitable to assess the accuracy of DFT+U. We demonstrate that tuning U to U=8 eV, which reproduces the experimental charge gap and magnetic moments (Fig. \ref{fig:dos}), brings phonon dispersions in agreement with experiments when a realistic AFM supercell of La$_\text{2}$CuO$_\text{4}$ [Fig. \ref{fig:cells} (b)] is used. We also find that the charge fluctuations induced by the breathing phonons near the Cu atoms are reduced at U=8 eV, consistent with the hardening of the bond-stretching phonons.

\begin{figure*}
\includegraphics[width=1\linewidth]{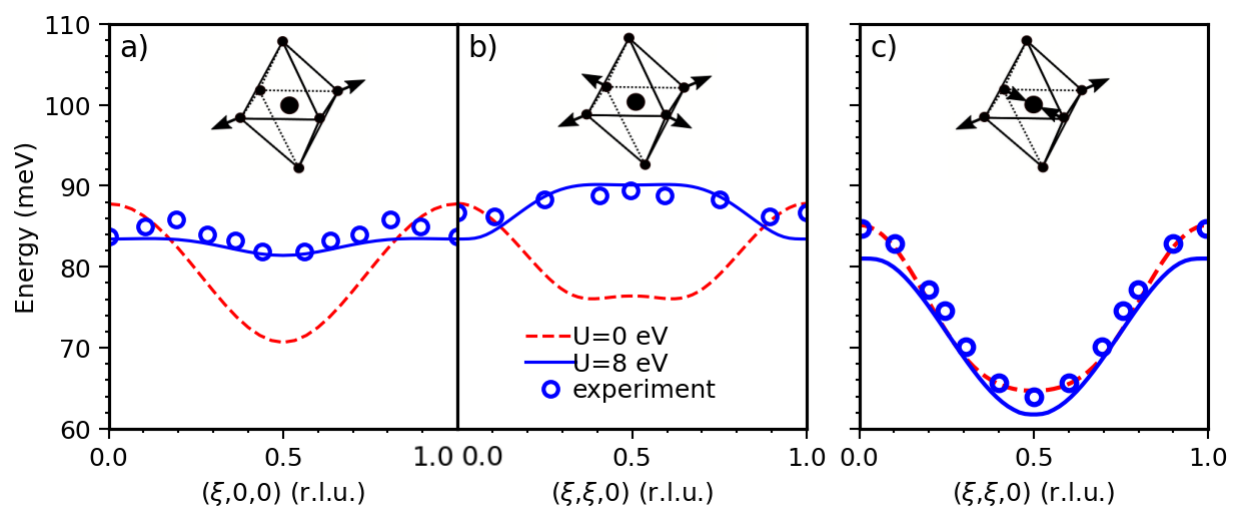}
        \caption{Bond-stretching phonon dispersions calculated using U=0 eV and U=8 eV. (a) Dispersion of the half-breathing branch along the $(\xi,0,0)$ reciprocal lattice direction. (b) Dispersion of the full-breathing branch along the $(\xi,\xi,0)$ reciprocal lattice direction. (c) Dispersion of the quadrupolar branch along the $(\xi,\xi,0)$ reciprocal lattice direction. The dashed red lines are the dispersions calculated with U=0 eV and the solid blue lines are dispersions calculated with U=8 eV. The reciprocal lattice directions are consistent with the HTT cell used in the experiments. The blue circles are experimental results from La$_\text{2}$CuO$_\text{4}$. The experimental data along $(\xi,0,0)$ are from Park $et~al.$ \cite{park2014evidence} and along $(\xi,\xi,0)$ are from Pintschovius $et~al.$ \cite{pintschovius2006oxygen}. The zone boundary eigenvectors of these modes are indicated by the diagrams in each plot.}
\label{fig:summary}
\end{figure*}

\section{methods}

We choose La$_\text{2}$CuO$_\text{4}$ for our study since the Cu-O bond-stretching phonons have been measured for all high-symmetry wave vectors and ranges of doping \cite{park2014evidence,pintschovius2006oxygen,pintschovius2005electron,miao2018incommensurate}. La$_\text{2}$CuO$_\text{4}$, like many perovskite-oxides, contains energy-lowering distortions that are static in the insulating phase resulting in a symmetry-lowered supercell \cite{zhang2020competing,furness2018accurate,boni1988lattice,birgeneau1987soft,billinge1996probing,bozin2015reconciliation,sapkota2021reinvestigation,zhang2020symmetry,varignon2019origin}. The structural phases of La$_\text{2}$CuO$_\text{4}$ have been explored using the meta-GGA SCAN functional \cite{sun2015strongly}, which also predicts gaps and moments in cuprates with accuracy comparable to DFT+U \cite{zhang2020competing,furness2018accurate}. It was shown that the insulating ground-state predicted for undoped La$_\text{2}$CuO$_\text{4}$ is further stabilized by including static lattice distortions, consistent with the observed soft modes \cite{boni1988lattice,birgeneau1987soft,billinge1996probing,bozin2015reconciliation,sapkota2021reinvestigation}. The distortions occur either in the form of Cu-O octahedral rotations resulting in the low-temperature-orthorhombic phase (LTO) or octahedral tilts resulting in the low-temperature-tetragonal (LTT) phase \cite{varignon2019origin,varignon2019mott,trimarchi2018polymorphous,zhang2020symmetry,lane2018antiferromagnetic,furness2018accurate}. Experiments show that the low-temperature structure of undoped La$_\text{2}$CuO$_\text{4}$ is LTO on average, but recent investigations showed that the local structure is likely fluctuating between the LTT and LTO phases \cite{lee2021hidden,bozin2015reconciliation,sapkota2021reinvestigation} and SCAN calculations showed that the energy difference between the LTO and LTT phases is comparable to computational errors \cite{furness2018accurate}. Nevertheless, lattice dynamics experiments use wave vectors from the high-temperature-tetragonal (HTT) structure \cite{pintschovius2006oxygen,park2014evidence,miao2018incommensurate,sapkota2021reinvestigation} [Fig. \ref{fig:cells} (a)]. Considering the apparent ambiguity in which distortions to include in the static structure, we chose the higher-symmetry LTT cell [Fig. \ref{fig:cells} (b)] over the LTO cell to simplify the analysis. 

For our DFT calculations, we used the projector augmented wave (PAW) method \cite{kresse1999ultrasoft,blochl1994projector} in the $\textsc{Vienna Ab-initio Simulation Package}$ ($\textsc{vasp}$) \cite{kresse1996efficiency,kresse1996efficient,kresse1993ab}. The standard PAW data sets included with $\textsc{vasp}$ were used to represent the core states of all atoms. All magnetism was assumed to be collinear and spin-orbit coupling effects were neglected. We choose DFT+U over SCAN. Both predict reasonably accurate electronic structures, but DFT+U allows us to tune the on-site repulsion by changing the U value. The U correction was applied to the Cu 3$d$ orbitals using the method proposed by Dudarev $et~al.$ where only a single parameter U$_{\text{eff}}\equiv$ U is used \cite{dudarev1998electron}. For exchange and correlation, we choose the LDA \cite{perdew1981self} (i.e. LDA+U) instead of GGA+U. Both the LDA+U and GGA+U methods have been successfully applied to electronic and structural properties of cuprates before \cite{zhang2007electron,pesant2011dft+,czyzyk1994local,anisimov1992spin,wei1994electronic,svane1992electronic,anisimov2004computation,elfimov2008theory,puggioni2009fermi,oh2011fermi}. Comparison between LDA+U and GGA+U in other transition metal oxides has shown that the physical properties are found to be similar, with the LDA+U usually requiring slightly larger values of U to reproduce experiments \cite{sun2008first,loschen2007first}. The results predicted here should not depend on the particular method used (i.e, LDA+U versus GGA+U), but the U value we find to be accurate should be compared to previous LDA+U studies. 

Relaxation and ground-state calculations were performed for integer values of U from 0 to 10 eV to determine the band gap and magnetic moments as a function of U. Note that U=0 eV corresponds to plain LDA. Phonons were calculated for a subset of the U values: U = 0, 1, 2, 5, 6, and 8 eV. 

In all calculations, the plane-wave energy cutoff was set to 650 eV and the total energy was required to converge to less than 1x10$^{-5}$ eV in the self-consistent cycle. During relaxation and ground-state calculations, Brillouin zone integrations were performed using a 12x12x6 $\Gamma$-centered k-point mesh and energy levels were smeared with a Gaussian function with width $\sigma$=0.01 eV to aid convergence. We chose a small $\sigma$ with a sufficiently dense k-point mesh to enable us to resolve small electronic gaps and magnetic moments as we increased U. Lattice parameters were fixed at the experimental values (a=5.360 $\text{\AA}$, c=13.236 $\text{\AA}$ \cite{cox1989structural}) for all U values and atomic positions were relaxed until the forces were less than 0.2 meV/$\text{\AA}$ on all atoms. To determine the electronic density of states, ground-state calculations were performed on the coarse 12x12x6 k-point mesh using the relaxed atomic positions followed by non-self-consistent calculations on a dense 24x24x12 k-point mesh integrated using the tetrahedron method \cite{blochl1994improved}. 

To calculate the phonon dispersions, we used the finite-difference approach in the code $\textsc{phonopy}$ \cite{phonopy} with 2x2x1 supercells for all U values. We included magnetic ordering when using symmetry to find the set of irreducible displacements. Each phonon calculation at a different U resulted in ground-state force calculations for 21 supercells with inequivalent atomic displacements frozen-in. We checked the convergence of the phonons from the 2x2x1 supercell against the results from a 3x3x1 supercell using U=6 eV. Note that the 2x2x1 supercell already contains 112 atoms, whereas the 3x3x1 supercell contains 252 atoms. Thus, calculations for the 3x3x1 supercell are very expensive and we could only test one case. We chose to test convergence using U=6 eV in anticipation of it predicting the insulating ground-state with reasonable electronic gap in La$_\text{2}$CuO$_\text{4}$. Since it is insulating, it should be poorly screened compared to the metallic ground-state and force constants should fall-off more slowly with distance. We found that the zone center and zone boundary energies of the longitudinal optical phonons considered here are nearly identical between the two supercell sizes. For some low-energy branches that are not smoothly dispersing, the dispersions are slightly modified across the Brillouin zone but the differences are not substantial and the zone-boundary and zone-center energies are nearly the same. For smoothly dispersing branches like the bond-stretching phonons, the 2x2x1 supercell is accurate. 

\begin{figure}
\includegraphics[width=1\linewidth]{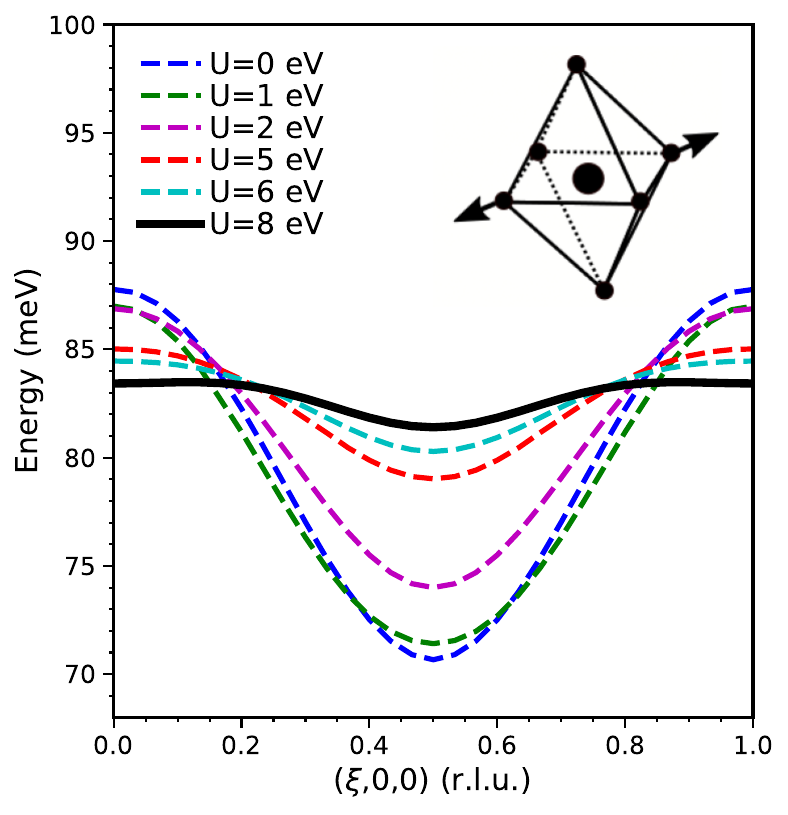}
        \caption{Half-breathing bond-stretching phonon calculated for the U values indicated in the figure. The zone boundary eigenvector is shown by the diagram in the plot. Only the branch that attained highest inelastic neutron scattering intensity in the $\bf{Q}$=$(4+\xi,0,0)$ zone is shown for each U.}
\label{fig:all_breathing}
\end{figure}

To analyze the effect of U on the charge density, we also calculated the charge density redistribution, $\Delta n = n-n^{ph}$, where $n$ is the self-consistent charge density in the unmodulated (i.e. fully relaxed) structure and $n^{ph}$ is the self-consistent charge density calculated with a phonon eigenvector frozen into the unit cell with a small amplitude. For small enough phonon amplitudes, $\Delta n$ shows the charge displaced due only to excitations of the valence electrons. To calculate $n^{ph}$, eigenvectors were identified by calculating the inelastic neutron structure factors in the correct zones (see below) and the atomic displacements of the eigenvectors were frozen in with the largest displacement away from equilibrium set to $\sim 0.09~\text{\AA}$. 

Since our calculations are based on the larger structurally distorted 28 atom cell of the low temperature AFM phases, the Brillouin zone is smaller with 84 branches that are closely spaced in energy with numerous anticrossings [Fig. \ref{fig:cells} (c)]. To relate the complicated dispersions in the LTT phase to the HTT cell, we calculated the inelastic neutron scattering structure factors S($\bf{Q}$,$\omega$) predicted by DFT+U in the reciprocal lattice units of the HTT cell. The color map in Fig. \ref{fig:cells} (c)  shows that only a few branches contribute to the scattering intensity in agreement with experiments. The intensity around 85 meV in Fig. \ref{fig:cells} (c) is from the half-breathing bond-stretching phonons. An alternative method, usually called unfolding, has been used to calculate effective band structures from supercell electron \cite{ku2010unfolding,popescu2010effective,popescu2012extracting} and phonon \cite{allen2013recovering,ikeda2018temperature,ikeda2017mode,samolyuk2021role,mu2020unfolding,kormann2017phonon} dispersions in the past. Our intuitive method (see the Supplemental Material for details \cite{supp}) benefits from direct comparability with experiments.

\section{results}

U=0 eV (i.e, plain LDA), gives the expected nonmagnetic, metallic ground-state. There are four electronic regimes for nonzero U: (i) AFM but metallic (U=1 eV), (ii) AFM and insulating but with unrealistically small gaps and moments (2$\geq$U$\geq$4 eV), (iii) AFM and insulating with reasonable gaps (5$\geq$U$\geq$8 eV), and (iv) AFM and insulating with unrealistically large gaps and moments (U$>$8 eV). The electronic charge gap opens with U=2 eV (Fig. \ref{fig:dos}) increasing with increasing U. U=6 eV gives gap/magnetic moments of 1.4 eV / 0.53 $\mu_\text{B}$, and U=8 eV gives 2.0 eV / 0.61 $\mu_{B}$m respectively. These values agree with experiments \cite{yamada1987effect,tranquada1988antiferromagnetism,vaknin1987antiferromagnetism,mitsuda1987confirmation,thio1990determination,uchida1991optical,cooper1990optical,kastner1998magnetic,ono2007strong} and previous calculations using DFT+U \cite{zhang2007electron,pesant2011dft+,czyzyk1994local,anisimov1992spin,wei1994electronic,svane1992electronic}. Gaps and moments that come out of U$\approx$8 eV have the best agreement with experiments. We found that for U values around U=8 eV, the electronic structure and lattice dynamics were not very sensitive to small changes in U. The electronic gaps and magnetic moments agreed reasonably well for the range of U values between 6 and 9 eV. There is considerable variability in the optical gap data in the literature, so we chose U=8 eV as the representative best value as it predicts magnetic moments closest to experiments when accounting for quantum fluctuations not present in the calculations. We note that the half-breathing LO phonon is actually in good agreement with experiments for both U=6 eV and U=8 eV. The agreement is similar for the full-breathing mode (which is sensitive to U) and the quadrupolar mode (which is not).

\begin{figure}
\includegraphics[width=1\linewidth]{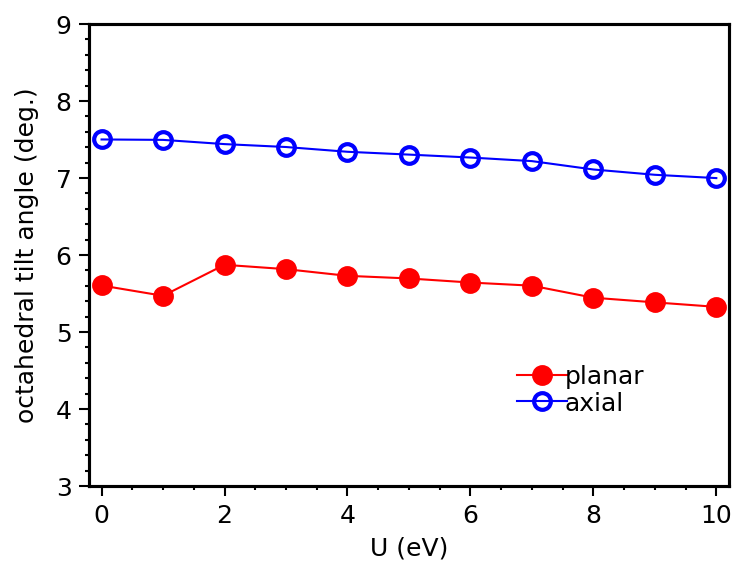}
        \caption{Octahedral tilt angles calculated from the $ab$ plane (planar) and from the $c$ axis (axial).}
\label{fig:angles}
\end{figure}

Two Cu-O planes in the HTT and LTT cells give two bond-stretching branches. Both are degenerate in the HTT phase, but the octahedral tilts lift the degeneracy in the LTT phase. In the case of the half-breathing phonon [Fig. \ref{fig:cells} (c)] the lower energy branch stretches bonds that are bent, whereas the higher energy branch stretches bonds that are straight. Lines in Fig. \ref{fig:summary} represent calculated dispersions of the bond-stretching phonons whose maximum intensity is in the Brillouin zone where experiments are performed. With U=8 eV, our dispersions are in striking agreement with experimental results. However the U=0 eV dispersions of the LO phonons [Figs. \ref{fig:summary}(a) and \ref{fig:summary}(b)] are significantly softer near the zone boundary than observed, consistent with previous DFT calculations \cite{cohen1990first,krakauer1993large,singh1996phonons,lebert2020doping,miao2018incommensurate,ahmadova2020phonon}. The calculated TO dispersion shown in (c) also agrees with experiments but is not affected by U. For U around 6 eV, the half-breathing phonon is weakly sensitive to small changes in U (Fig. \ref{fig:all_breathing}). The trend is nearly identical for the full-breathing branch. The improved agreement of the bond-stretching phonons in insulating La$_\text{2}$CuO$_\text{4}$ relative to the metallic U=0 eV ground state is consistent with a series of microscopic model calculations based on linear response theory where the charge response was seperated into local and nonlocal parts \cite{falter1997origin,falter1993effect,falter1995phonon,falter2000effect,falter2001nonlocal,falter2002influence,falter2006modeling,bauer2009impact}.

To check if the change in the bond-stretching phonon energies is an artifact of \emph{internal} relaxation of the unitcell (i.e. the atomic positions) at fixed lattice parameters, we checked the tilt angle of the Cu-O octahedra in the relaxed structure at each U (Fig. \ref{fig:angles}). For all U values used in this paper, the octahedral rotations we calculated agree well with previous SCAN, GGA, and LSDA calculations (4.3-8.5$^{\circ}$) \cite{furness2018accurate,lane2018antiferromagnetic} which all over estimate the experimental tilt angles ($\sim$3.5$^{\circ}$) \cite{cox1989structural,yamada1989spin}. Moreover, aside from the small blip around U=1 eV, the octahedral tilt angles evolve smoothly with only very small $\sim$ 5\% change in the relaxed octahedral tilt angles across all U values. The behavior at U=0 and U=1 eV is due to the inability of LDA with small U to predict a stable insulating, tilted ground-state. From U=2 eV through U=8 eV, the bond-stretching phonon energies harden drastically (Fig. \ref{fig:all_breathing}) while the tilt angles are nearly unchanged. Thus, it is unlikely that the hardening of the bond-stretching phonons is due to a modulation of the Cu bonds imposed by relaxing the structure at fixed lattice parameters. At U=8 eV, the tilt angles are close to previous calculated values and the bond-stretching phonon energies are in excellent agreement with experiment. 

We note that since the calculations tend to over estimate the tilt angles, the $\sim$2 meV splitting of the two no-longer degenerate bond-stretching branches in Fig. \ref{fig:cells} is over estimated too. The small tilt angles observed in experiments would result in smaller splitting between these branches and, with finite experimental resolution combined with the intrinsic line widths ($\sim$1 meV \cite{park2014evidence}) of the bond-stretching phonons, it is unlikely that such small splitting could be resolved.

\begin{figure}
\includegraphics[width=1\linewidth]{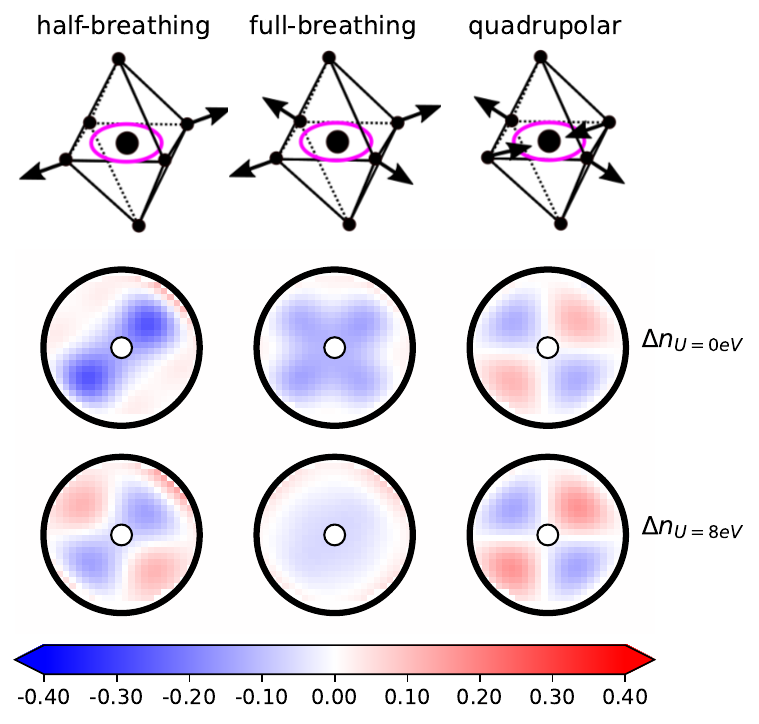}
        \caption{Comparison of the amount of charge pumped by the zone boundary phonons, $\Delta n$ (defined in the text), into the vicinity of the Cu atoms where the on-site potential is applied. Colorbar indicates the amount of charge pumped into/out-of a small volume around the Cu atoms, in units of $e/\text{\AA}^3$. The top and bottom rows of colormaps show the amount of charge displaced by the phonons calculated with U=0 eV (LDA) and U=8 eV, respectively. The displacement pattern of the Cu-O octahedra for each phonon is shown in the top row; the magenta circle shows the region around the Cu atom plotted in the color maps. The subscript on the labels on the right indicates U used to calculate $\Delta n$ in that row.}
\label{fig:chg_den_Cu}
\end{figure}

\section{discussion}

In the bond-stretching phonons, the motion of the in-plane O atoms modulates the charge density in the Cu-O bonds, pumping charge into/out of the vicinity of the Cu atoms. The amount of charge displaced by the bond-stretching phonons at the zone boundary is presented in Fig. \ref{fig:chg_den_Cu}. The color maps show the excess charge, $\Delta n$, induced by bond-stretching atomic lattice displacements in a small region around the Cu atoms where the on-site potential is applied. Dark-blue regions are positions in the unit cell where a (relatively) large amount of charge is depleted by the phonon; dark red regions are positions where a large amount of charge is added. We show the charge modulation calculated for the phase of the phonon displacement shown in the figure; if the phase were rotated by $\pi$, the O atoms moving away from the Cu sites would instead be moving towards them and the sign of the charge modulation would flip. However, the analysis below would be the same.

For the LO bond-stretching phonons, the amount of the charge that is pumped depends on U because U sets the energy cost to modulate the charge around the Cu atoms. If U is small, there is relatively little energy cost to pump charge into/out of the vicinity of the Cu atoms. On the other hand, with U=8 eV, there is a substantial energy cost to change the amount of charge near the Cu atoms. The half- and full-breathing phonons depend strongly on U since the motion of the in-plane O atoms (top row in Fig. \ref{fig:chg_den_Cu}) changes the volume of the Cu octahedra, displacing the charge. Our calculations bear this out: There is a considerably larger amount charge displaced into/out of the vicinity of Cu atoms by the LO phonons with U=0 eV (middle row in Fig. \ref{fig:chg_den_Cu}) compared with U=8 eV (bottom row in Fig. \ref{fig:chg_den_Cu}). On the other hand, the TO quadrupolar mode induces nearly the same charge modulation with U=0 eV and U=8 eV as discussed below. We also calculated the charge modulation using U=2 eV and the results are very similar to U=0 eV (see the Supplemental Material \cite{supp}). 

The phonon energy depends on the amount of screening, i.e, the energy is proportional to the amplitude of electronic charge fluctuations driven by atomic vibrations. Charges are free to redistribute between the Cu and O orbitals when U=0, but increasing U blocks these fluctuations for the half-breathing and breathing modes, and the  amount of screening is reduced. As a result, these modes harden with increasing U.

For the quadrupolar mode, the motion of two in-plane O atoms outward is compensated by the inward motion of the other two in-plane O atoms, so the eigenvector does not substantially modulate the volume of the octahedra. Note how charges depleted in two of the lobes due to this quadrupolar displacement in Fig. \ref{fig:chg_den_Cu} are compensated by increased charge density in the other two, so the net occupation of the Cu site does not change. This is the reason that it is not sensitive to the value of U. A similar argument is true for the bond-stretching phonons near the zone center, whose energies also do not depend on U. We note that it was previously found that the symmetry of the zone boundary quadrupolar mode prohibits coupling to the Cu 3$d$ orbitals \cite{falter2002influence}. On the other hand, the LO phonons are not prohibited from coupling to Cu 3$d$ orbitals, consistent with their dependence on U \cite{giustino2008small}. 

In a nutshell, DFT+U points at the following mechanism: the hardening of the LO bond-stretching phonons with U can be understood by considering the charge density redistribution induced by the atomic displacements. With U=8 eV, modulating the charge on the Cu atom is unfavorable due to the imposed on-site repulsion. Rather than be compressed into the Cu-O octahedra, the charge in the Cu-O bonds tends to delocalize. In undoped La$_\text{2}$CuO$_\text{4}$, the delocalized charge wants to go into the Cu 3$d$ orbitals on other sites, but transferring charge into those orbitals also costs an energy that scales with U. The charge around the Cu atoms becomes rigid (Fig. \ref{fig:chg_den_Cu}, bottom row) and the modulated charge has to be excited across the electronic gap (which also scales with U). The LO bond-stretching phonons are coupled to the Cu 3d orbitals, so their energies increase. On the other hand, when holes are introduced into La$_\text{2-x}$Sr$_\text{x}$CuO$_\text{4}$ by doping, there are low-energy charge excitations that do not require changing the occupation of the Cu 3$d$ orbitals \cite{uchida1991optical}. The charge fluctuations induced by the LO bond-stretching phonons are no longer frustrated by the on-site repulsion and the LO zone boundary phonons are soft \cite{park2014evidence,pintschovius2006oxygen}.  

For small U values, the agreement with the overdoped experimental results might lead one to form analogies between doping and varying U. In fact, it has been assumed that LDA or GGA calculations of undoped cuprates represent the overdoped compounds and the resulting bond-stretching dispersions do usually agree with experiments on overdoped cuprates \cite{krakauer1993large,miao2018incommensurate,giustino2008small,lebert2020doping}. We caution that agreement between the LDA or GGA dispersions and the overdoped compound is merely coincidental. Electron-phonon coupling quantities like line widths and spectral functions in hole-doped La$_\text{2-x}$Sr$_\text{x}$CuO$_\text{4}$ are not reproduced by GGA \cite{giustino2008small,reznik2008photoemission}. A recent many-body perturbation theory calculation starting from DFT+U wave functions accurately reproduced the observed electronic spectral function and showed that the density of states at the Fermi level was different from GGA calculations \cite{li2021unmasking}. It has already been established that electronic-structures from explicitly hole-doped DFT+U calculations are qualitatively different than undoped LDA or GGA electronic-structures \cite{anisimov2004computation,elfimov2008theory,puggioni2009fermi,oh2011fermi} and quantities like electron-phonon line widths, which result from integrals over the Fermi surface, will be qualitatively different too. In future work, we intend to validate this concept by calculating electron-phonon properties in cuprates using doped DFT+U ground-states. 

\section{conclusion}

To summarize, we calculated the phonon spectrum of undoped La$_\text{2}$CuO$_\text{4}$ using the LTT supercell with the energy-lowering distortions that are present in the real material and the DFT+U method with U=8 eV. Our calculations reproduced the experimental band gap and antiferromagnetic moments. The calculated bond-stretching dispersions computed with U=8 eV agreed with experiments, demonstrating the sensitivity of the LO bond-stretching dispersions to the on-site repulsion, U, which frustrates modulating the charge around the Cu atoms. This is consistent with the experimental result that the LO bond-stretching branches soften with hole-doping, since doping permits low-energy charge excitations that do not require pumping charge onto the Cu atoms. These results should be valid for transition-metal oxides in general. We showed that the DFT+U method combined with the correct supercell is a robust framework for modeling phonons in the undoped cuprates and other perovskite oxides with complex, interrelated structural and electronic degrees of freedom. 

\section{acknowledgments}

The authors would like to thank J.M. Tranquada, A. Holder, and I.I. Mazin for helpful discussions and suggestions. This work was supported by the DOE, Office of Basic Energy Sciences, Office of Science, under Contract No. DE-SC0006939. Portions of this research were performed using the Eagle computer operated by the Department of Energy's Office of Energy Efficiency and Renewable Energy and located at the National Renewable Energy Laboratory, and the RMACC Summit supercomputer which is supported by the National Science Foundation (Awards No. ACI-1532235 and No. ACI-1532236), the University of Colorado Boulder, and Colorado State University.

\bibliography{ref}

\end{document}